\documentclass[fleqn,usenatbib,useAMS,usegraphicx]{rasti}

\usepackage{newtxtext,newtxmath}

\usepackage[T1]{fontenc}

\DeclareRobustCommand{\VAN}[3]{#2}
\let\VANthebibliography\thebibliography
\def\thebibliography{\DeclareRobustCommand{\VAN}[3]{##3}\VANthebibliography}

\usepackage{graphicx}	
\usepackage{amsmath}	
\usepackage{longtable}
\usepackage{tabularx}
\usepackage[inkscapelatex=false]{svg}
\usepackage{pifont}
\newcommand{\cmark}{\ding{51}}%
\newcommand{\xmark}{\ding{55}}%
\usepackage{subcaption}
\usepackage{tikz}
\usetikzlibrary{arrows.meta,positioning}

\tikzset{
  layer/.style={draw,rounded corners,minimum width=2.8cm,minimum height=0.9cm,align=center},
  nodefeat/.style={draw,rounded corners,minimum width=2.2cm,minimum height=0.8cm,align=center},
  arrow/.style={-Stealth,thick},
  gnode/.style={circle,draw,minimum size=6pt,inner sep=1pt},
  gedge/.style={-},
  attnedge/.style={-{Stealth[length=2mm]},dashed},
}

\title[GNNs for Cosmic Web Classification]{Learning the Cosmic Web: Graph-based Classification of Simulated Galaxies by their Dark Matter Environments}

\author[D. Kololgi et al.]{
D. Kololgi,$^{1}$\thanks{E-mail: dakshesh.kololgi.23@ucl.ac.uk (DK)}
K. Naidoo,$^{2,3}$
A. Saintonge,$^{4,1}$
O. Lahav$^{1}$
\\
$^{1}$Department of Physics and Astronomy, University College London, Gower Street, London WC1E 6BT, UK\\
$^{2}$Institute of Cosmology and Gravitation, University of Portsmouth, Burnaby Road, Portsmouth, PO1 3FX, UK\\
$^{3}$Max-Planck-Institut f\"{u}r Astronomie, K\"{o}nigstuhl 17, 69117 Heidelberg, Germany\\
$^{4}$Max-Planck-Institut f\"{u}r Radioastronomie, Auf dem H\"{u}gel 69, 53121 Bonn, Germany
}

\date{Accepted XXX. Received YYY; in original form ZZZ}

\pubyear{\the\year{}}

\begin{document}
\label{firstpage}
\pagerange{\pageref{firstpage}--\pageref{lastpage}}
\maketitle

\begin{abstract}

We present a novel graph-based machine learning classifier for identifying the dark matter cosmic web environments of galaxies. Large galaxy surveys offer comprehensive statistical views of how galaxy properties are shaped by large-scale structure, but this requires robust classifications of galaxies' cosmic web environments. Using stellar mass-selected IllustrisTNG-300 galaxies, we apply a three-stage, simulation-based framework to link galaxies to the total (mainly dark) underlying matter distribution. Here, we apply the following three steps: First, we assign the positions of simulated galaxies to a void, wall, filament, or cluster environment using the T-Web classification of the underlying matter distribution. Second, we construct a Delaunay triangulation of the galaxy distribution to summarise the local geometric structure with ten graph metrics for each galaxy. Third, we train a graph attention network (GAT) on each galaxy's graph metrics to predict its cosmic web environment. For galaxies with stellar mass $\mathrm{>10^9 M_{\odot}}$, our GAT+ model achieves an accuracy of $85\,\%$, outperforming graph-agnostic multilayer perceptrons and graph convolutional networks. Our results demonstrate that graph-based representations of galaxy positions provide a powerful and physically meaningful way to infer dark matter environments. We plan to apply this simulation-based graph modelling to investigate how the properties of observed galaxies from the Dark Energy Spectroscopic Instrument (DESI) survey are influenced by their dark matter environments.

\end{abstract}




\section{Introduction}
\label{sec:introduction}

Understanding the role of galaxy environments is essential for constraining models of galaxy evolution. Galaxies and their host dark matter haloes form and evolve within the cosmic web - the large-scale organisation of dark matter into void, wall, filament, and cluster environments. Since the cosmic web is fundamentally a feature of the dark matter distribution, its environments provide the dynamical framework within which galaxies grow. A growing body of evidence shows that these environments influence galaxy properties beyond what is captured by halo mass alone \citep{white_core_1978}, for example, through anisotropic mass accretion, transport, and preprocessing \citep{cautun_evolution_2014, kraljic_impact_2020}. Furthermore, recent studies have found that filaments can efficiently deliver cold gas into cluster environments, providing fuel for massive central galaxies \citep{ma_neutraluniversemachine_2024, martin_multi-filament_2019}. Large-scale structures provide substantial local density information, which ultimately determines the properties of dark matter halos hosting galaxies \citep{green_tidal_2019, buncher_probabilistic_2020}. To study how galaxy properties depend on their environment and how this may affect their evolution, it is therefore necessary to first obtain a robust classification of the underlying dark-matter cosmic web environments.

Various methods, as outlined by \citet{libeskind_tracing_2018}, have been developed to define the cosmic web in dark matter, reflecting the different tracers available in galaxy surveys. In simulations, where the full dark matter field is accessible, classifications often rely on dynamical information to provide physically motivated definitions of cosmic web environments such as the eigenvalues of the Hessian or velocity shear tensors \citep{hahn_properties_2007, foreroromero_dynamical_2009} or multi-scale morphology filters that identify filaments and walls across a range of smoothing scales \citep{aragon-calvo_multiscale_2010, cautun_nexus_2013}. In contrast, galaxy surveys consist of biased and incomplete tracers, and methods that are applied directly to galaxy distributions are uniquely designed to study different aspects of the cosmic web. For example, \citet{sousbie_disperse_2013} measures the robustness of topological features to define filaments and other structures. Others include smoothed density-field reconstructions \citep{alonso_recovering_2016}; graph-based friends-of-friends or minimum spanning tree approaches \citep{alpaslan_large-scale_2014}; and Bayesian inference of the underlying matter field (e.g. BORG; \citealt{jasche_bayesian_2013}).

We can train machine learning models on cosmological simulations, where dark matter environments are unambiguously defined, producing robust classifiers for galaxy surveys to study galaxy evolution. Machine learning classifiers are scalable and generalisable due to their adaptable architectures and training schemes. In particular, graphs can be combined with machine learning algorithms to provide an efficient and scalable framework to infer large-scale structures without reconstructing a density field. However, survey systematics and selection functions that, for example, arise from magnitude limitations and selection functions impact the graph construction. These biases can be forward-modelled by leveraging the large volumes of synthetic data available today in simulations and mock catalogues. A less computationally intensive approach to this problem has recently been developed by \citet{naidoo_methods_2025} for cosmological parameter inference, which involves assigning probabilities to galaxies based on their survey selection functions. 

In this paper, we present the development of a graph-based machine learning classifier for galaxy environments in the dark matter cosmic web. The machine learning models we present are trained on the IllustrisTNG-$300$  simulation \citep{nelson_illustristng_2021} with the true cosmic web environments assigned by the Hessian method \citep{hahn_properties_2007} using the T-Web classification method \citep{foreroromero_dynamical_2009}. In \autoref{sec:Methods}, we describe how cosmic web environment labels are extracted from the Hessian eigenvalues of the dark matter density field. This is followed by an introduction to the IllustrisTNG simulations and the graph constructions we tested. Finally, we describe a baseline neural network architecture based on a multi-layer perceptron (MLP) \citep{tsizh_large-scale_2020}, and then graph neural networks (GNNs) based on graph convolution \citep{kipf_semi-supervised_2017} and graph attention \citep{velickovic_graph_2018,brody_how_2022} architectures, respectively. Then, \autoref{sec:Results} compares the performance of our baseline MLP and GNN architectures with alternative machine learning classifiers. We also outline the architectural and training adaptations that were used to prevent overfitting and maximise generalisability. Finally, in \autoref{sec:Discussion}, we discuss the interpretability of our graph metrics and the performance of our model. Then we discuss the physical effects and survey systematics that must be overcome to apply our model to large-scale galaxy surveys.

\section{Methods}
\label{sec:Methods}

\subsection{Density field-based cosmic web classification}

The cosmic web lacks a universally agreed-upon definition; its characterisation varies significantly between studies depending on their scientific goals, methodological choices, and the nature of available data \citep{libeskind_tracing_2018}. For this reason, the choice of classification scheme applied to the IllustrisTNG simulation is critical, as it defines the physical assumptions (inductive biases) that our machine learning model will learn and use to generalise.

\citet{libeskind_tracing_2018} and  \citet{buncher_probabilistic_2020} outline several classification methods for the cosmic web. A comprehensive classification scheme capable of resolving all four cosmic web environments—voids, walls, filaments, and clusters—is essential to validate our graph-based machine learning framework. After preliminary testing of multiple classification schemes, we chose the T-Web scheme \citep{foreroromero_dynamical_2009} that uses the sign of the Hessian eigenvalues of the gravitational potential to classify each point on a discrete simulation grid.

Classically, the cosmic web arises naturally from the Zeldovich approximation \citep{zeldovich_gravitational_1970}, which describes structure formation as a consequence of anisotropic gravitational collapse. In this linear regime, the displacement of matter is proportional to the gradient of the gravitational potential.

In the Zeldovich formalism, the position of a fluid element $\mathbfit{x}$ at time $t$ (Eulerian coordinate) is related to its initial position $\mathbfit{q}$ (Lagrangian coordinate) given by
\begin{equation}
    \mathbfit{x}(\mathbfit{q},t) = \mathbfit{q} + D(t)\mathbfit{s}(\mathbfit{q}),
    \label{eq:zelmapping}
\end{equation}
where $D(t)$ is the linear growth function and $\mathbfit{s}(\mathbfit{q})\,=\,-\nabla_{q}\Phi(\mathbfit{q})$ represents the initial displacement field, defined as the gradient of the gravitational potential field over the initial Lagrangian coordinates.

The formation of non-linear structures predicted by the Zeldovich approximation emerges from the non-linear evolution of density encoded in the Jacobian of the coordinate transformation between Lagrangian and Eulerian space, despite $\mathbfit{s}(\mathbfit{q})$ being linear. The Jacobian derives directly from \autoref{eq:zelmapping} to the left-hand side of \autoref{eq:zeljacobian},

\begin{equation}
    \frac{\partial x_i}{\partial q_j} = \delta_{ij} + D(t)\Psi_{ij},
    \label{eq:zeljacobian}
\end{equation}

where $\Psi_{ij} = \frac{\partial s_i}{\partial q_j} = -\frac{\partial^2 \Phi (\mathbfit{q})}{\partial q_i \partial q_j}$, defined as the deformation tensor of the scalar gravitational potential field evaluated in Lagrangian space. The eigenvalues of this Jacobian describe the local deformation of a volume element (curvature) and determine whether it collapses into a cluster (collapse along 3 axes), filament (collapse along 2 axes), wall (collapse along 1 axis) or expands into a void (no collapse) \citep{doroshkevich_spatial_1970}. T-Web identifies cosmic web environments by similarly computing the eigenvalues of the Hessian of the gravitational potential,

\begin{equation}
    H_{i,j}(\bar{\mathbfit{x}})\,=\,\frac{\partial^2 \phi}{\partial x_{i} \partial x_{j}}\Bigg|_{\bar{\mathbfit{x}}},
    \label{eq:hessianpotential}
\end{equation}

where $i,\,j\,=\,1,2,3$, denote the three spatial dimensions. The signs of the three Hessian eigenvalues are calculated for each point. Three positive eigenvalues correspond to a matter collapse along all three axes, and therefore a cluster; two positive eigenvalues to a filament; one positive eigenvalue to a wall. No positive eigenvalues correspond to an expansion in all three axes and are therefore classified as a void. Using the sign of the eigenvalues represents an eigenvalue threshold value of $\mathrm{\lambda_{th} = 0}$, an idealisation that assumes that matter collapse has fully completed.

However, \citet{foreroromero_dynamical_2009} argues for an eigenvalue threshold $\lambda_{\mathrm{th}} > 0$. In the Zeldovich approximation, the magnitude of each positive eigenvalue is proportional to the rate of collapse along its axis, $\mathrm{t_{collapse}}\propto \frac{1}{\lambda}$. Therefore, if $\lambda \approx 0$, then the collapse timescale can exceed the Hubble time, meaning that no significant structure would have formed in the age of the Universe. Although dynamical considerations justify why $\mathrm{\lambda_{th}} > 0$ is necessary \citep{foreroromero_dynamical_2009}, the exact choice of the threshold is decided empirically. They find that already at $\mathrm{\lambda_{th} = 0.1}$, the resulting classification accurately reproduces the visual impression of the cosmic web, but voids percolate unrealistically. After testing in different environments and smoothing scales, a threshold value between $0.2 \lesssim \lambda_{\mathrm{th}} \lesssim 0.4$ yielded robust classifications and distinct under-dense voids.

\subsection{IllustrisTNG-300 simulations}

In this study, we apply this formalism to the particle field of the IllustrisTNG-300 simulations. To do this, we apply a Gaussian filter with scale $\mathrm{R_{s} = 2\,\text{h}^{-1}\,Mpc}$ to smooth the field and remove sensitivity to features on smaller scales and systematics from the mass assignment scheme used to compute the density field from dark matter particles. The number of Hessian eigenvalues that exceed the threshold value determines whether the cosmic web environment is a void (zero eigenvalues above the threshold), wall (one), filament (two), or cluster (three). Separately, we create a mock catalogue of subhalo particles with a stellar mass cut, $\mathrm{M_c}$, to ensure that they are galaxy proxies. Upon classifying each grid-point within the simulation cube using T-Web, the coordinates of each mock galaxy are used to associate it to its environment in the density field.

\begin{table*}
    \centering
    \begin{tabularx}{\textwidth}{|l|X X |X X |X X |X X|}
        \hline
         & \multicolumn{4}{c|}{$10^{9}\,\mathrm{M}_{\odot}$} & \multicolumn{4}{c|}{$10^{10}\,\mathrm{M}_{\odot}$} \\ 
        \hline
         \textbf{Environment} & \multicolumn{2}{c|}{Unbuffered} & \multicolumn{2}{c|}{Buffered} & \multicolumn{2}{c|}{Unbuffered} & \multicolumn{2}{c|}{Buffered} \\ 
        \hline
        Void     & $10,773$ & $3.79\%$ & $8313$ & $3.63\%$ & $2009$  & $2.07\%$  & $1531$ & $1.95\%$ \\ 
        Wall     & $69,019$ & $24.3\%$ & $55,293$ & $24.1\%$ & $20,242$  & $20.8\%$ & $16,139$ & $20.6\%$ \\ 
        Filament & $128,025$ & $45.1\%$ & $103,362$ & $45.1\%$ & $46,746$ & $48.1\%$ & $37,778$ & $48.1\%$ \\ 
        Cluster  & $76,152$ & $26.8\%$ & $62,282$ & $27.2\%$ & $28,236$ & $29.0\%$  & $23,056$ & $29.4\%$ \\ \hline
        \textbf{Total}    & $283,969$ & $100\%$ & $229,250$ & $100\%$ & $97,233$ & $100\%$ & $78,504$ & $100\%$ \\ \hline
    \end{tabularx}
    \caption{Galaxy counts in each cosmic (T-Web) environment with mass-cuts of $\mathrm{10^{9}\,M_{\odot}}$ and $\mathrm{10^{10}\,M_{\odot}}$. Buffered columns show how the distribution of galaxy counts change after ignoring a $10\,\mathrm{Mpc}$ region from each face of the simulation cube due to edge effects. It should be noted that galaxies in the buffer region are included in the graph construction but only excluded from training, validation and testing of the machine learning models. Model development was conducted with a mass cut of $\mathrm{10^{10}\,M_{\odot}}$ due to the shorter training time however all results are presented with a mass cut of $\mathrm{10^{9}\,M_{\odot}}$.}

    \label{tab:galaxy_counts_props}
\end{table*}

\begin{figure*}
    \centering
    \includegraphics[width=\linewidth]{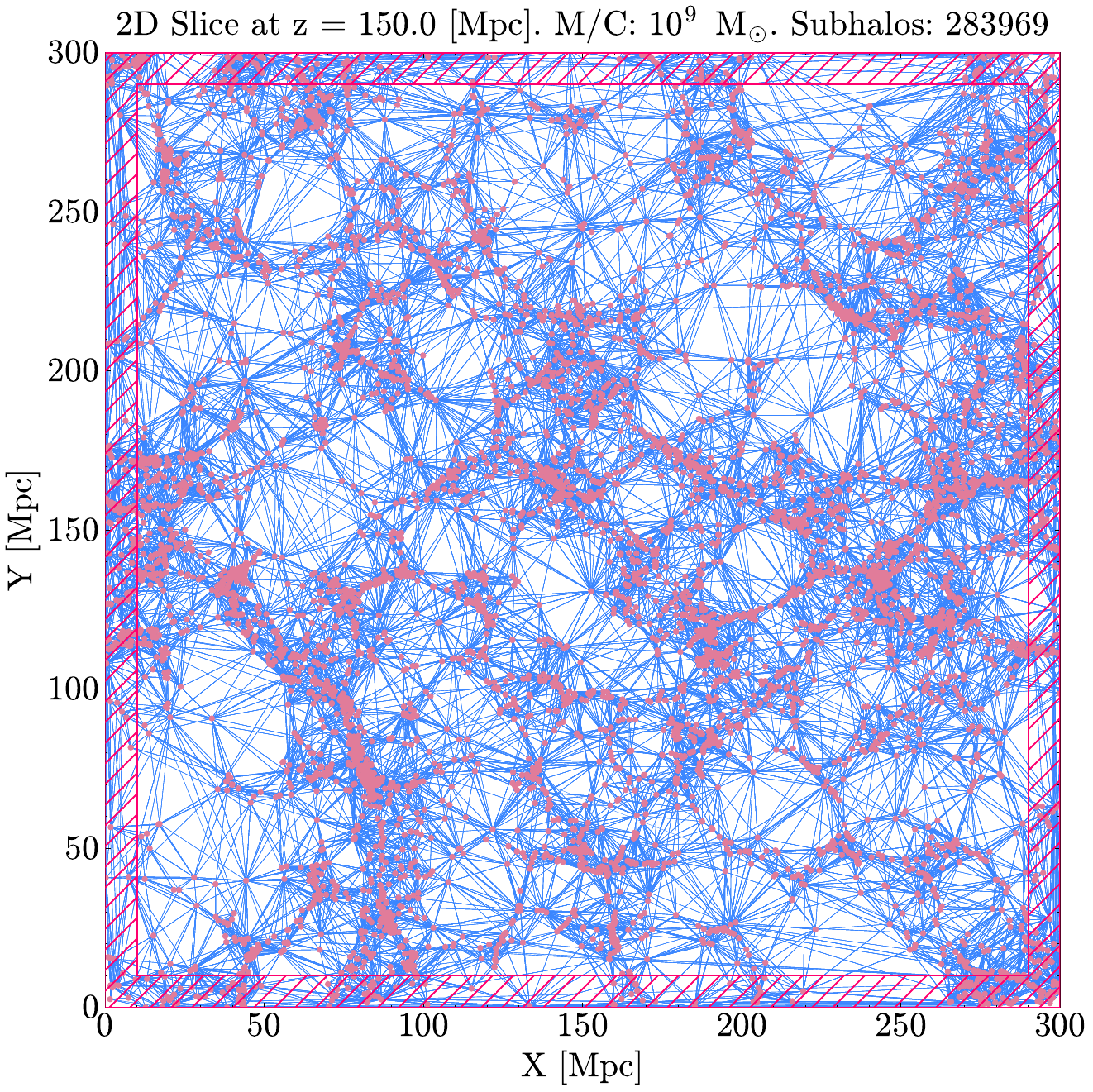}
    \caption{Two-dimensional slice of the Delaunay triangulation constructed from the spatial distribution of IllustrisTNG galaxies. The slice is $1\,\mathrm{Mpc}$ thick in the X-Y plane at $Z = 150\,\mathrm{Mpc}$, and shows the network of edges connecting neighbouring galaxies above a stellar mass cut of $10^{9}\,M_{\odot}$. The Delaunay graph captures the underlying geometric structure of the galaxy distribution, naturally tracing voids, walls, filaments, and clusters within the cosmic web. The $10\,\mathrm{Mpc}$-thick hatched region around the box is the buffer region. Galaxies in this region have unphysical graph metrics due to the spuriously high density of edges resulting from the simulation boundaries.}
    \label{fig:2d_delaunay_tng_10_9}
\end{figure*}

\begin{figure}
    \centering
    \includegraphics[width=\linewidth]{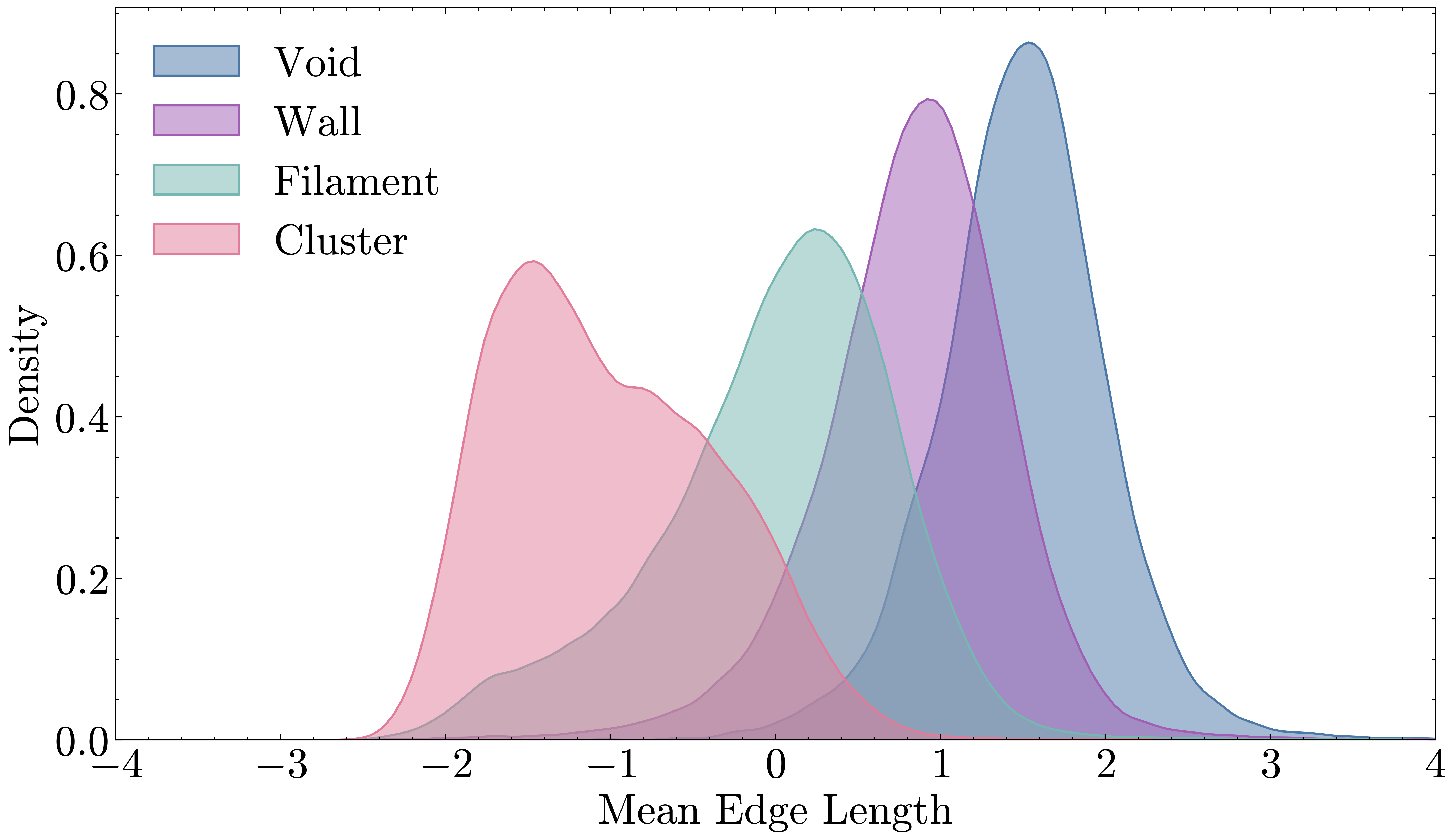}
    \caption{Distributions of mean edge lengths for each IllustrisTNG galaxy (above a $\mathrm{10^9 \,M_{\odot}}$ stellar mass cut). The edges are determined by the Delaunay triangulation graph. The mean edge lengths are scaled using a Box-Cox power transform, as are all the graph metrics, to make them more Gaussian-like for faster training and better numerical stability. This example demonstrates that even a single graph metric can partially discriminate the four cosmic web environments, motivating the use of a combination of graph metrics to capture richer structural information.}
    \label{fig:meanel}
\end{figure}

For the T-Web classification, we follow \citet{foreroromero_dynamical_2009} for an eigenvalue threshold value $\mathrm{\lambda_{th} = 0.2}$ and a smoothing scale $R_s = 2\,\mathrm{h}^{-1}\,\mathrm{Mpc}$. The grid spacing was maintained at $\Delta x,\Delta y, \Delta z = \frac{205\,\mathrm{h}^{-1}\,\mathrm{Mpc}}{512} \, \approx 0.4 \mathrm{h}^{-1}\,\mathrm{Mpc}$, which means $512^{3}$ boxes within the $300\, \mathrm{Mpc}$ ($205\, \mathrm{h}^{-1}\,\mathrm{Mpc}$) simulation cube, where the cosmic web environment is classified. The smoothing scales above are valid as long as $R_s > \Delta x, \Delta y, \Delta z$. Initially, we also tested the NEXUS+ algorithm \citep{cautun_nexus_2013}; however, it produced a significant class imbalance, resulting in unrealistic cosmic web reconstructions and poor performance of the ML classifiers. Henceforth, we focus on the results from the T-Web classification. \autoref{tab:galaxy_counts_props} outlines the distribution of galaxies in each environment with a mass cut of $\mathrm{M_c >10^{9}\, M}_{\odot}$. This mass cut was chosen to reflect the likely mass range of DESI survey galaxies \citep{zou_photometric_2019}. 

\subsection{Mapping galaxies to graphs}

The second step of our method is to construct a graph from the positions of the IllustrisTNG galaxies to summarise the local geometric structure into quantitative metrics. As a motivating example, \autoref{fig:meanel} demonstrates the distributions of mean edge lengths from a Delaunay triangulation applied to the IllustrisTNG galaxies. We see a clear separation of the mean edge lengths of each cosmic web environment, indicating that multiple such graph metrics can be used to build a representation of cosmic web environments based solely on the relative positions of galaxies. 

We defined up to ten graph metrics that summarised (non-exhaustively) the amount of geometric information at the node level (where each node is a galaxy). For an expanded definition, please refer to \autoref{appendix:graph_metric_defn}. We initially explored three types of graphs: minimum spanning trees \citep[MSTs;][]{kruskal_Shortest_1956a, prim_shortest_1957,naidoo_beyond_2020}, complex graphs \citep{de_regt_network_2018, tsizh_large-scale_2020}, and Delaunay triangulations \citep{delone_boris_nikolaevich_b_1934}. Minimum spanning trees connect each node once without cycles to minimise the total edge length. Complex graphs are constructed as a nearest-neighbour graph based on the radial distance. For our testing, we followed \citet{tsizh_large-scale_2020} and \citet{de_regt_network_2018} in setting the radius of neighbours $l = 2\,\mathrm{h^{-1}\,Mpc}$. However, this comes with increased computational cost when the number of nodes is very large. Finally, the Delaunay graph is constructed by doing a Delaunay triangulation of the galaxies. A Delaunay triangulation for a set of points (galaxy positions) in a plane (or in higher dimensions) subdivides their convex hull into triangles (tetrahedra in three-dimensions or simplex more generally) whose circumcircles do not contain any of the points \citep{delone_boris_nikolaevich_b_1934}.

\begin{table*}
    \centering
    \begin{tabular}{|c|c|c|c|c|}
        \hline
        & \textbf{MST} & \textbf{Complex graph} & \textbf{Delaunay graph} & \textbf{Description} \\ \hline
        \textbf{degree} & \cmark & \cmark & \cmark & Connectivity \\ \hline
        \textbf{average degree} & \cmark & \cmark & \xmark & Connectivity \\ \hline
        \textbf{katz centrality} & \cmark & \xmark & \xmark & Centrality \\ \hline
        \textbf{degree centrality} & \cmark & \cmark & \xmark & Centrality \\ \hline
        \textbf{eigenvector centrality} & \cmark & \cmark & \xmark & Centrality \\ \hline
        \textbf{constraint} & \cmark & \xmark & \xmark & Investment \\ \hline
        \textbf{mean edge length} & \cmark & \xmark & \cmark & Geometry \\ \hline
        \textbf{mean neighbour edge length} & \cmark & \xmark &  \xmark & Geometry\\ \hline
        \textbf{mean neighbour's neighbour edge length} & \cmark & \xmark & \xmark & Geometry \\ \hline
        \textbf{clustering} & \xmark & \cmark & \cmark & Clustering \\ \hline
        \textbf{min edge length} & \xmark & \xmark & \cmark & Geometry \\ \hline
        \textbf{max edge length} & \xmark & \xmark & \cmark & Geometry \\ \hline
        \textbf{tetrahedral density} & \xmark & \xmark & \cmark & Density \\ \hline
        \textbf{neighbour tetrahedral density} & \xmark & \xmark & \cmark & Density \\ \hline
        \textbf{$\mathrm{I}_1$} & \xmark & \xmark & \cmark & Local Morphology \\ \hline
        \textbf{$\mathrm{I}_2$} & \xmark & \xmark & \cmark & Local Morphology \\ \hline
        \textbf{$\mathrm{I}_3$}  & \xmark & \xmark & \cmark & Local Morphology \\ \hline
        \textbf{\# Features} & 8 & 5 & 10 & - \\ \hline
    \end{tabular}

    \caption{Comparison of the graph metrics we tested across the MST, complex graph, and Delaunay graph. The degree, average degree, clustering and centrality metrics are inspired by \citet{tsizh_large-scale_2020}. The degree, average degree, centrality measures, clustering, and constraint metrics were calculated using methods within the \texttt{networkx} \citep{hagberg_exploring_2008} package and all other graph metrics were derived manually using the definitions in \autoref{appendix:graph_metric_defn}. While we tested the MST and Complex graphs, we found that the Delaunay graph produced the highest accuracy with the least computational costs. For this reason, we focus on results from the Delaunay graph in this paper.}
    \label{tab:graph_metrics}
\end{table*}

Testing revealed that the Delaunay triangulation graph was the most performant for classifying galaxies. This is likely because the MST is a subset of the Delaunay graph \citep{grami_discrete_2023}, with fewer connections and containing less geometric and topological information. \citet{libeskind_tracing_2018} found that MST-based methods were useful for finding filaments but that a secondary MST graph was required to detect structures on smaller scales \citep{alpaslan_large-scale_2014, alpaslan_galaxy_2014}. Secondly, the complex graph could not compete with the performance of the Delaunay graph due to the computational limitations of the linking length. While the radius $l = 2\,\mathrm{h}^{-1}\,\mathrm{Mpc}$ was used to match with \citet{tsizh_large-scale_2020}, it could not classify galaxies in all four cosmic web environments. Graphs with larger radii were not feasible given the available computational power. A longer radius may have captured structures on larger scales, such as filaments, walls, and voids in galaxies. Thus, the Delaunay graph was the most effective at capturing all four cosmic web environments.

The Delaunay triangulation graph was implemented with the \texttt{scipy.spatial.Delaunay}\footnotemark[1] class that can take two- or three-dimensional coordinates as inputs. In our method, we use the three-dimensional coordinates of the IllustrisTNG galaxies. The \texttt{Delaunay} class contains the \texttt{simplices} method which outputs the indices of the vertices of all simplices in the triangulation, as shown in \autoref{fig:2d_delaunay_tng_10_9} (2D slice at $Z = 150\,\mathrm{Mpc}$). The list of simplices is then used to construct a \texttt{networkx}\footnotemark[2] graph, which is used to calculate a set of graph metrics.

\footnotetext[1]{\url{https://docs.scipy.org/doc/scipy/reference/generated/scipy.spatial.Delaunay.html/}}
\footnotetext[2]{\url{https://networkx.org/documentation/stable/index.html/}}

\subsection{Graph-derived metrics}
\label{sec:graph_derived_metrics}

Each galaxy, represented by a graph node, is assigned ten metrics, derived from the Delaunay graph's connectivity, which serve as feature inputs to the neural network models. Table~\ref{tab:graph_metrics} summarises the refined set of graph metrics adopted in this work, based on \citet{hong_network_2015}, \citet{de_regt_network_2018}, and \citet{tsizh_large-scale_2020}. In addition to previously used topological quantities, we introduced measures of edge length, density, and shape eigenvalues to probe both local density and morphological anisotropy. The definitions of these ten metrics are provided in \autoref{appendix:graph_metric_defn} and summarised in \autoref{tab:graph_metrics}. We describe the model architecture in further detail in \autoref{sec:model_architecture}. Henceforth, we refer to the features as graph metrics. Physically, the graph metrics describe how galaxies are connected within the underlying dark matter, tracing the density, clustering, and spatial arrangement of matter that defines the cosmic web.

\begin{figure}
    \centering
    \includegraphics[width=\linewidth]{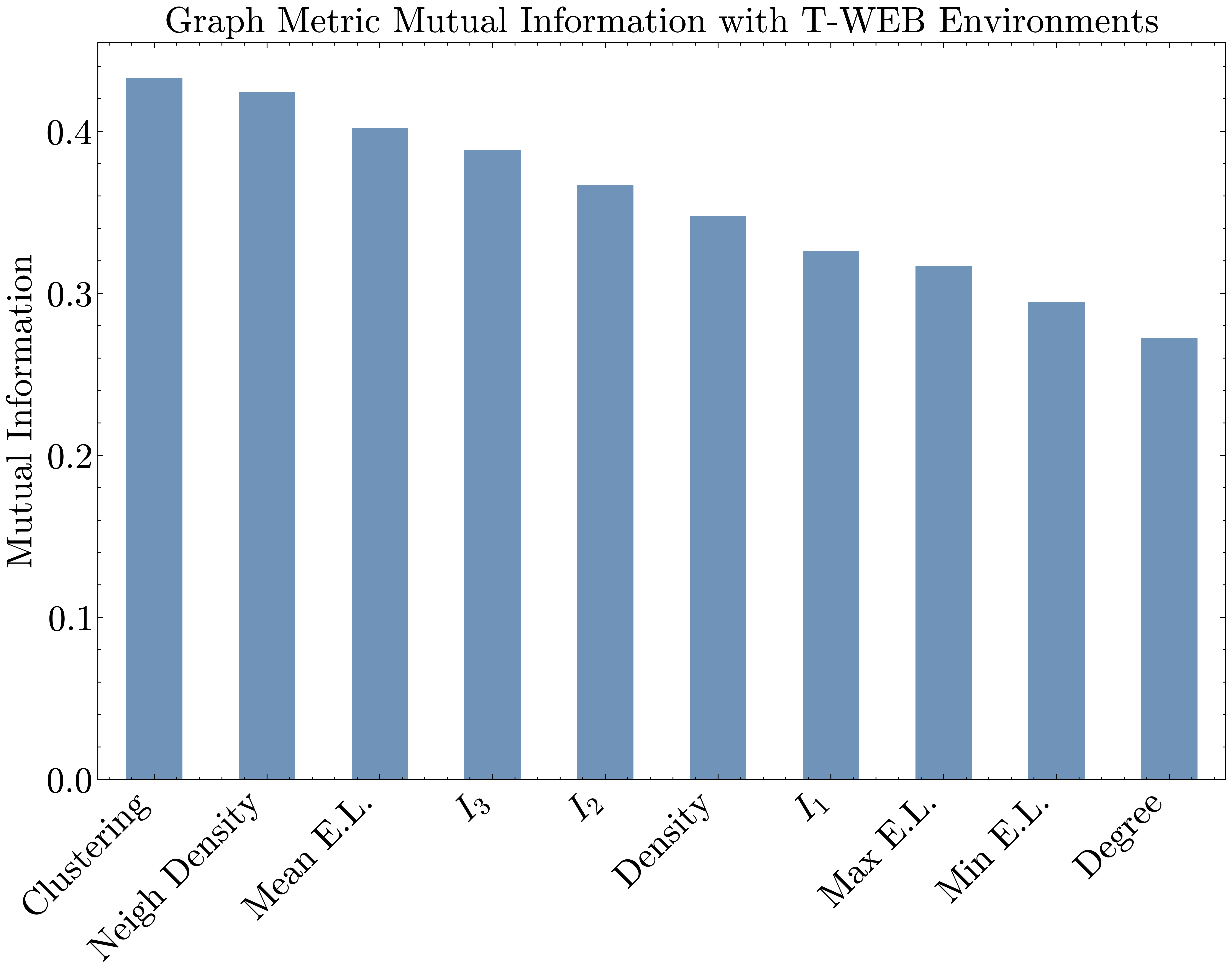}
    \caption{The mutual information \citep{ross_mutual_2014, kraskov_estimating_2004} quantifies the statistical dependency between each node-level graph metric and the categorical environments (void, wall, filament, cluster) defined by the T-WEB classifier. Mutual information captures both non-linear and non-monotonic relationships. The clustering coefficient and neighbour density exhibit the strongest association with cosmic web environments, indicating their higher discriminative power for capturing relevant local structures. At the same time, degree and minimum edge length show relatively weaker dependencies.}
    \label{fig:MI}
\end{figure}

To ensure that only the most informative and non-redundant graph metrics are used, we used the mutual information to quantify the dependence of each metric on the cosmic web environments. Metrics with low mutual information or high redundancy were discarded, which yielded the final set of Delaunay graph metrics listed in \autoref{tab:graph_metrics} and their mutual information with the cosmic web environments is shown in \autoref{fig:MI}.

Furthermore, the boundaries around the simulation cube force the Delaunay algorithm to connect galaxies in this region with long and thin `sliver' tetrahedra, biasing the graph metrics for those galaxies. As a result, we introduced a $10\,\mathrm{Mpc}$ wide buffer region along each face of the simulation cube (see \autoref{fig:2d_delaunay_tng_10_9}) to remove these edge effects in the 10 graph metrics. \autoref{tab:galaxy_counts_props} shows that there were 78,504 (out of 97,233) viable galaxies for $\mathrm{M_c =} 10^{10}\,\mathrm{M}_{\odot}$ and 229,250 (out of 283,969) galaxies for $\mathrm{M_c} = 10^{9}\,\mathrm{M}_{\odot}$.

\subsection{Machine learning architectures for learning the cosmic web environments from graph metrics}
\label{sec:model_architecture}

In this section, we outline the third step in our method that involves predicting the cosmic web environments of each galaxy based on its graph metric features after learning the mapping between these metrics and the T-Web dark matter environments. We describe the architectural choices for the baseline multiplayer perceptron (MLP) model before reviewing the fundamentals of graph neural networks (GNNs) and how neural networks can be built on graphs using the principles of graph representation learning. For a more comprehensive introduction to graph neural networks and graph representation learning, refer to \citet{bronstein_geometric_2021, kipf_semi-supervised_2017, velickovic_graph_2018}.

\subsubsection{Multi-layer perceptron network}

An MLP is a feed-forward neural network consisting of fully connected layers that apply successive transformations to the input data. With non-linear activations, MLPs are universal function approximators \citep[see][]{hornik_multilayer_1989}, making them powerful tools for capturing the complex relationships between galaxy graph metrics and cosmic web environments. For a more comprehensive introduction to common deep learning practices please refer to \citet{lecun_deep_2015}.

Each layer performs a linear transformation of the input graph metric feature vector,

\begin{equation}
    \mathbfit{x}_{in} \, \mathbfss{W}^{T} + \mathbfit{b} = \mathbfit{x}_{out},
    \label{eq:mlp}
\end{equation}

where $\mathbfss{W}$ are the connection weights and $\mathbfit{b}$ the biases. The non-linear ReLU activation function, $\mathrm{ReLU(\mathbfit{x}) = \max(0, \mathbfit{x})}$, is applied after each hidden layer.

\begin{figure*}
    \centering
    \includegraphics[width=\linewidth]{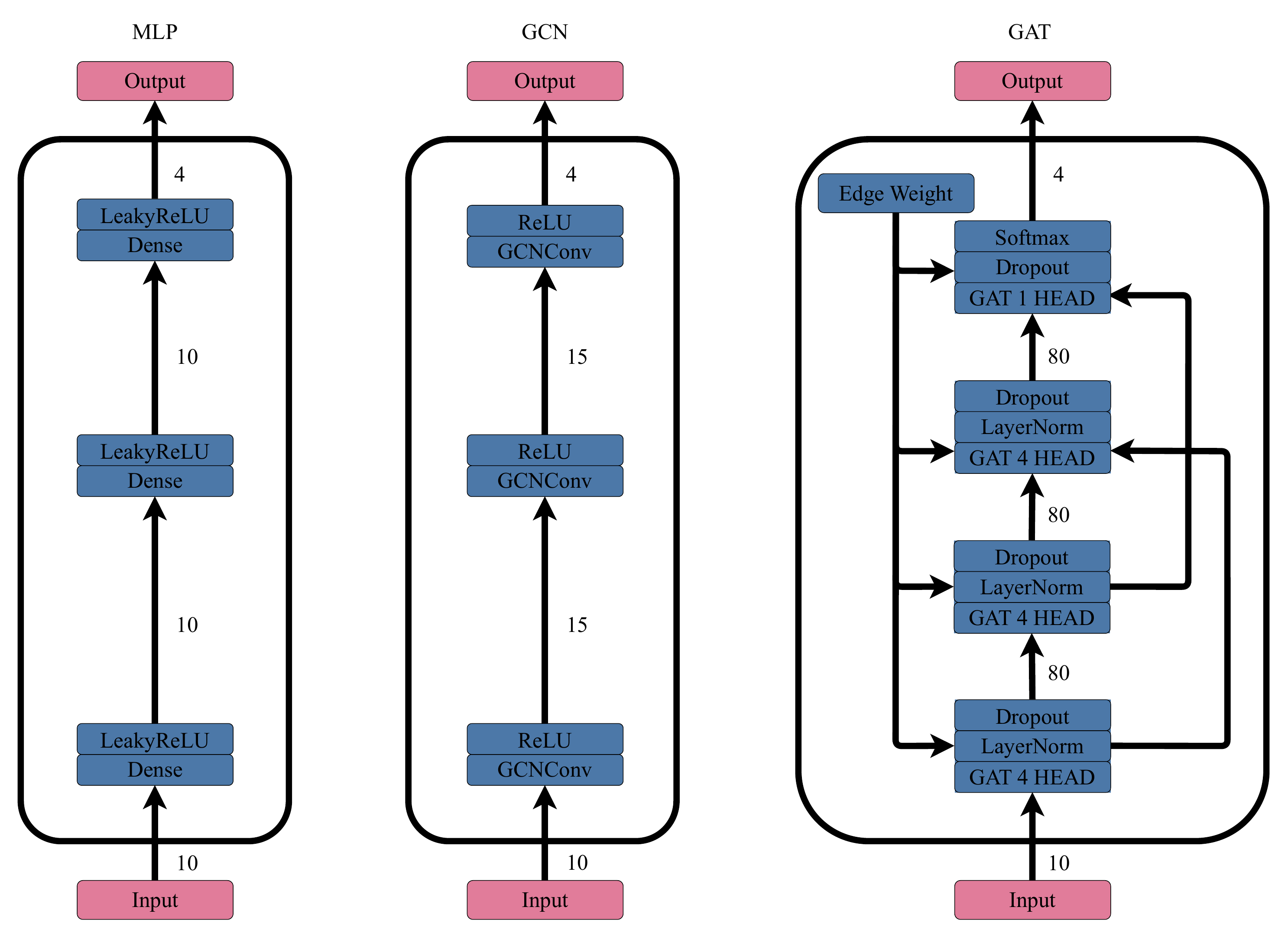}

    \caption{Overview of the baseline and graph-based neural network architectures explored in this work.
The MLP (left) serves as the baseline model, taking node features as independent inputs passed through successive fully-connected layers. The GCN (middle) introduces relational inductive biases by aggregating information from connected nodes in the Delaunay graph, enabling feature propagation along edges. The GAT+ (right) extends this by applying multi-head attention mechanisms and edge features, allowing the network to learn the relative importance of neighbouring nodes. The architectures were refined through iterative experimentation, adjusting the number of layers, hidden dimensions, and normalisation or dropout configurations until convergence performance and stability were optimised across validation runs.}
    \label{fig:mlpgnngatplusarchitecture}
\end{figure*}

Our MLP architecture (shown in \autoref{fig:mlpgnngatplusarchitecture}) consists of three layers: an input layer of 10 neurons (corresponding to the 10 graph metrics), one hidden layer of 10 neurons, and an output layer of 4 neurons representing the four cosmic web environments. The number of neurons in each layer determines the amount of data compression of the input graph metrics. The final layer applies a Softmax activation to convert the outputs into probabilities that sum to unity, with the highest probability defining the predicted environment.

The model is trained using the cross-entropy loss function and the ADAM optimiser \citep[see][]{kingma_adam_2017} with a learning rate of $\mathrm{\eta} = 0.001$. The dataset is split into $64\%$ training, $16\%$ validation, and $20\%$ test sets. A batch size of $\mathrm{B} = 16$ was found to yield stable convergence, balancing computational efficiency and gradient smoothness. Training was carried out over multiple epochs of the training dataset, with the validation losses monitored to prevent overfitting. The model was implemented in \texttt{PyTorch} \citep{ansel_pytorch_2024}.

\subsubsection{Graph neural network preliminaries}
\label{sec:graph_neural_network_preliminaries}

While the MLP model serves as a useful baseline for cosmic web classification, it treats each galaxy's graph metrics as independent inputs, ignoring the relational structure encoded by the Delaunay graph. In contrast, GNNs embed this structure directly into the model architecture by using a message passing scheme that aggregates information from neighbouring galaxies at each layer. This allows the network to capture how local connectivity patterns relate to the cosmic web environment. By exploiting relational information, GNNs can be more robust and generalisable than MLPs when applied to new datasets \citep[see][]{battaglia_relational_2018}.

A graph of $\mathrm{N}$ galaxies has no intrinsic ordering. The Delaunay triangulation defines relationships (edges) but not a specific ordering. For a GNN model to learn structures independent of the galaxy ordering, a GNN must be insensitive to how galaxies are indexed. This property, known as permutation equivariance, ensures that shuffling the node order in the input permutes the output. For example, given a function $\mathrm{f}$, that takes as input a graph in matrix form, $\mathbfss{A}$, and a permutation of the node orderings given by a matrix, $\mathbfss{P}$, then,

\begin{equation}
    \mathrm{f(\mathbfss{P}\mathbfss{X}}) = \mathrm{\mathbfss{P}^{T}\mathrm{f}(\mathbfss{X})\mathbfss{P}},
    \label{eq:permutationequivariance}
\end{equation}

defines permutation equivariance. Any architecture with this property recognises the same structures (e.g. filaments) regardless of ordering, enabling generalisation to unseen graphs of galaxies. This symmetry encodes a physically meaningful inductive bias: the Universe does not distinguish between galaxy labels, only the gravitational relationships represented as edges matter. In practice, GNN layers achieve permutation equivariance through permutation-invariant aggregations of neighbouring node features for a given galaxy, followed by node updates. For a detailed description, please see \citet{bronstein_geometric_2021}.

The Delaunay graph (or the other constructions) can be represented as an undirected graph (\autoref{eq:graphdef}). 

\begin{equation}
    \mathrm{\mathcal{G} = \left( V, E \right),\, V = \{ \nu_1, \nu_2,...,\nu_N \},\,E \subseteq V \times V}
    \label{eq:graphdef}
\end{equation}

$V$ is the set of vertices with each element $\nu_i$ corresponding to a galaxy. $E$ is the set of edges between galaxies. Here $N$ is the number of galaxies. Given $\mathcal{G}$, we can represent the edges of the graph with an adjacency matrix $\mathbfss{A}$. For undirected graphs, the adjacency matrix is a symmetric $N\times N$ matrix with elements $A_{ij}$,

\begin{equation}
    \mathbfss{A}_{ij} = \begin{cases}
        1\, (\text{$w_{ij}$ for weighted graphs})\,\text{if}\, (\nu_i,\nu_j) \in E, \\
        0 \, \text{otherwise.}
    \end{cases}
    \label{eq:adjacencyelement}
\end{equation}

Each node in the graph has a feature vector, $\mathbfit{x}_i$, consisting of the metrics described in \autoref{tab:graph_metrics}. Stacking these gives the \textit{node feature matrix} $\mathbfss{X} = \{\mathbfit{x}_1 , \mathbfit{x}_2,...,\mathbfit{x}_i,..., \mathbfit{x}_N\}$ ($\mathbfss{X} \in \mathbb{R}^{N \times 10}$). Feature representations at deeper layers are denoted $\mathbfss{H} = \{\mathbfit{h}_1, \mathbfit{h}_2,...,\mathbfit{h}_i,...,\mathbfit{h}_N\}$.

\subsubsection{Graph convolution network}
\label{sec:graph_convolution_network}

The first GNN architecture we use is a graph convolutional network (GCN) \citep{kipf_semi-supervised_2017, defferrard_convolutional_2017, wu_simplifying_2019}. Our implementation of the GCN derives from \citet{kipf_semi-supervised_2017} and is given by,

\begin{equation}
    \mathbfss{H}^{(k+1)} = \sigma \left( \tilde{\mathbfss{D}}^{-\frac{1}{2}} \tilde{\mathbfss{A}}   \tilde{\mathbfss{D}}^{-\frac{1}{2}} \mathbfss{H}^{(k)} \mathbfss{W}^{(k)} \right).
    \label{eq:generalgcnlayermatrix}
\end{equation}

We represent the update in a layer $(k+1)$ for the graph metric features as a matrix transformation of the adjacency matrix $\mathbfss{A}$, which handles the aggregation of neighbour features across the entire graph. Following \citet{kipf_semi-supervised_2017}, we employ self-connections $\tilde{\mathbfss{A}} = \mathbfss{A} + \mathbfss{I}_{\mathrm{N\times N}}$, so that each node also propagates its own features forward. Without this, the node's feature representation after one layer would depend only on its neighbours. Secondly, they also apply symmetric degree normalisation $\tilde{\mathbfss{D}}^{-\frac{1}{2}} \tilde{\mathbfss{A}} \tilde{\mathbfss{D}}^{-\frac{1}{2}}$ using the degree matrix $\tilde{\mathbfss{D}} \in \mathbb{R}^{\mathrm{N\times N}}$, of $\tilde{\mathbfss{A}}$ with $\tilde{D}_{ii} = \sum_j \tilde{A}_{ij}$, which encodes the degree of each node. This rescales neighbour contributions, weighting the aggregation so that high-degree nodes (e.g. cluster galaxies) do not dominate low-degree nodes (e.g. void galaxies) and improves numerical stability during training. 

Each layer contains a learnable weight matrix, $\mathbfss{W}^{(k)}$, that linearly combines the input features to produce a new latent representation. For example, in the first layer, it could be a linear combination of degree, clustering, and mean edge length. The same $\mathbfss{W}^{(k)}$ is applied to every node, ensuring permutation equivariance and allowing the GCN model to interpret a galaxy's graph metrics in the context of its neighbours. For example, learning that a galaxy with a high degree and a high local density indicates a cluster environment.

The number of neurons in each layer, $F_{(k)}$ is a hyperparameter that controls the compression of input features. Then the node feature matrix has dimensions $\mathrm{\mathbfss{X} \in \mathbb{R}^{N \times F(0)} = \mathbb{R}^{N \times 10}}$. We found that three GCN layers with $15$ neurons in each hidden layer achieved the highest test accuracy with minimal complexity (see \autoref{fig:mlpgnngatplusarchitecture}). The dimensions of the weight matrices for each layer are then $\mathrm{\mathbfss{W} \in \mathbb{R}^{F(0) \times F(1)} = \mathbb{R}^{10 \times 15}}$, $\mathrm{\mathbfss{W} \in \mathbb{R}^{F(1) \times F(2)} = \mathbb{R}^{15 \times 15}}$, $\mathrm{\mathbfss{W} \in \mathbb{R}^{F(2) \times F(3)} = \mathbb{R}^{15 \times 4}}$. Adding more layers or neurons did not improve performance.

\subsubsection{Graph Attention Network}
\label{sec:graph_attention_network}

Graph attention networks (GATs) \citep{velickovic_graph_2018, monti_fake_2019, zhang_efficient_2020} are a more expressive flavour of GNN compared to GCNs. While GCNs aggregate neighbour features using fixed, degree-normalised weights, GATs extend this approach by assigning learnable, feature-dependent weights to each graph edge through an attention mechanism \citep{bahdanau_neural_2014}. This enables the model to weigh the importance of different neighbours when aggregating features during training. These learnable attention coefficients are a function of the feature representations of two graph nodes $\alpha^{(k)}_{i,j} = \alpha^{(k)}(\mathbfit{h}_i, \mathbfit{h}_j)$ and can be packaged into an adjacency-like matrix $\pmb{\alpha}^{(k)} \in \mathbb{R}^{N\times N}$ for the entire graph. The implementation of the GAT model used in this investigation is a slight modification of the original \citep{velickovic_graph_2018} version developed by \citet{brody_how_2022} for dynamic attention. Henceforth, any reference to GAT will refer only to the modified version shown in matrix form in \autoref{eq:generalgatlayermatrix}. To obtain the graph-wide update for each layer, the learnable weight matrix for the features $\mathbfss{W}^{(k)}_{msg}$ transforms the features of the previous layer $\mathbfss{H}^{(k)}$; which is in turn transformed by the attention coefficient matrix $\pmb{\alpha}^{(k)}$ that aggregates neighbour node features depending on the learnt attention coefficients. 

In each GAT layer, the attention coefficients are computed independently of the feature-mixing transformation $\mathbfss{W}^{(k)}_{msg}$. Separately, a scoring function is applied to each pair of linearly projected node features to assess the relative importance of each neighbour. These raw scores are normalised with a softmax function over each node's neighbourhood to produce the final attention weights, yielding a data-dependent aggregation scheme that adapts to the graph structure and features during training. For further details, see \citet{brody_how_2022}.

\begin{equation}
    \mathbfss{H}^{(k+1)} = \sigma \left( \pmb{\alpha}^{(k)} \mathbfss{H}^{(k)} \mathbfss{W}^{(k)}_{\text{msg}} \right) 
    \label{eq:generalgatlayermatrix}
\end{equation}

So far, we have assumed a single attention head, corresponding to a single scalar coefficient per graph edge. However, in our implementation, we use four attention heads to increase the expressiveness and stability of our GAT model. Each head independently computes its own set of attention coefficients and feature-mixing weights, allowing the model to capture different relational signals in parallel. For example, one head may prioritise nodes with similar structural features, while another may emphasise particular connectivity patterns or densities. \citet{brody_how_2022} demonstrated how multi-headed attention helps to disentangle this structural information to produce more informative node representations compared to a single-headed GAT. The outputs of these heads are then concatenated or averaged to form a richer and more robust representation that reduces the risk of any single attention pattern dominating. This is particularly useful for our purposes, as we have a heterogeneous graph with sharp boundaries between cosmic web environments. For an in-depth derivation of how the attention coefficients in the single- and multi-headed cases are calculated, see \citet{brody_how_2022}.

\section{Results}
\label{sec:Results}

\subsection{Loss and accuracy}
\label{sec:loss_and_accuracy}

\begin{figure*}
    \centering
    \includegraphics[width=\textwidth]{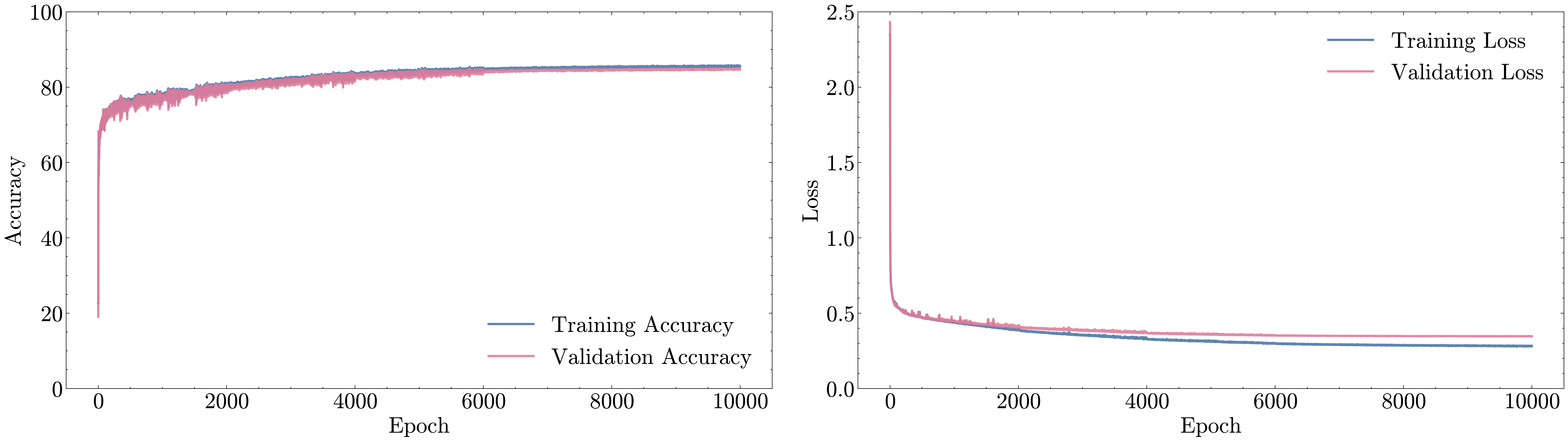}
    \caption{Training and validation performance of the GAT+ model. Accuracy (left) and loss (right) curves are shown for the training and validation datasets over 10,000 epochs for the $\mathrm{10^{9}\,M_{\odot}}$ stellar mass cut. The convergence and close overlap between training and validation curves indicate stable optimisation and minimal overfitting.}
    \label{fig:gat_4H_plus_10_9_loss_accuracy_curves}
\end{figure*}

Here, we describe and justify our choices for training the MLP, GCN, and GAT models. For all three models, $70\%$ of the galaxies in \autoref{tab:galaxy_counts_props} were randomly assigned to the training set for the models, $21\%$ as validation galaxies during training and $9\%$ as test galaxies after the training process was completed. Each sub-set of the data was stratified to maintain the same proportions of each cosmic web environment as in the unbuffered sample (\autoref{tab:galaxy_counts_props}). The training process for the GAT+ model with a mass-cut of $\mathrm{10^{9}\,M_{\odot}}$ is described in \autoref{fig:gat_4H_plus_10_9_loss_accuracy_curves}. A weight $w_{\mathrm{class}}$ was added to the cross-entropy loss function for each cosmic web environment (see \autoref{tab:environmentweights}) to account for class imbalances in the training data. (Please refer to the second sub-column in \autoref{tab:galaxy_counts_props} to see the differences in proportion of galaxies in each cosmic web environment). \autoref{eq:weights} shows that the weights were based on the number of galaxies in each cosmic web environment, $n_{\mathrm{galaxies\,in\,class}}$, the total number of galaxies, $N_{\mathrm{samples}}$, and the number of cosmic web environments, $n_{\mathrm{classes}}$. 

\begin{equation}
    w_{\mathrm{class}}\,=\,\frac{N_{\mathrm{galaxies}}}{n_{\mathrm{classes}}\times n_{\mathrm{galaxies\, in\, class}}}.
    \label{eq:weights}
\end{equation}

Since the training data are imbalanced, an unweighted loss function would bias the model towards predicting the majority, which minimises the average error. Therefore, applying weights to the loss function is valid under the assumption that the proportions of void, wall, filament, and cluster galaxies reflect the real Universe. \autoref{fig:gat_4H_plus_10_9_loss_accuracy_curves} also shows the convergence to a minimum loss and maximum accuracy. The closeness of the loss and accuracy curves for the training and validation datasets indicates that the model generalises well to unseen data and is not overfitting.

Typically, ML algorithms do not process one data point (galaxy in this case) at a time; instead, they process a stack of data by leveraging an additional matrix dimension that represents the batch size or number of galaxies being processed simultaneously. We used batch processing while training the baseline MLP model. However, this was not feasible for the GNN models. After testing a range of batch size values to optimise test accuracy, we found the optimal value for the batch size to be $B = 16$. When $B=8$ (the lower limit of our testing range), the loss updates for each epoch became noisy and took longer to converge. This is likely because smaller batches estimate the loss function gradients with a higher degree of variance, meaning that the optimiser takes longer to find a clear direction for downward loss. In particular, we found this to affect the accuracy of the MLP model in classifying void galaxies. In contrast, when $B=32$, the MLP model generally converged to a higher loss and lower accuracy during training, and consistently performed $\sim5\%$ worse on the test data set. This is likely because the loss updates per iteration were too big and the optimiser fluctuated around local loss minima.

On the other hand, for the GCN and GAT models, we implemented full-batch training, utilising the entire graph in each training epoch. However, we maintain training, validation, and testing sets of nodes. The message passing component of GNNs means that the features of neighbouring nodes that might not be in the training dataset could be aggregated. This is an example of transductive learning \citep{schwartz_machine_2024}, represented by the adjacency and degree matrices in \autoref{eq:generalgcnlayermatrix}.

The loss and accuracy convergences were noisier for the GCN and GAT models due to the recursive neighbourhood aggregation (message-passing functionality) between layers, posing additional challenges compared to the MLP model. For example, messages from high-degree nodes might amplify through the network and cause gradient explosions or oscillating updates. For the GCN model, the adjacency matrix that describes the graph structure is degree-normalised (see \autoref{eq:adjacencyelement}) precisely to prevent exploding values due to repeated multiplication over multiple layers \citep{kipf_semi-supervised_2017}. 

The fixed neighbour averaging of the GCN model favours homophilous graphs with no connections between nodes of different target classes \citep{zhu_beyond_2020, zhu_graph_2021}. This is the case for our work because the Delaunay triangulation is a heterophilous graph, as it ensures connections between galaxies whose T-Web environments are different (for example, between clusters and walls). \citet{ma_is_2023} empirically found that GCNs can perform well on heterophilous graphs with adjustments, and \citet{cai_graphnorm_2021} and \citet{kipf_semi-supervised_2017} outlined that careful hyperparameter tuning, using dropout layers (or other regularisation procedures) and using attention mechanisms help GCN models disentangle signals from misleading neighbours. This motivated our use of the GAT model as it models the neighbour influence more  precisely.

We found that introducing a single dropout layer with a dropout probability of $50\%$ (as \citet{kipf_semi-supervised_2017} found) in the intermediate layers (see \autoref{fig:mlpgnngatplusarchitecture}) acts to remove periodic noise from the loss and accuracy curves while training. Additional regularisation procedures were required to stabilise the training process for the GAT model. Normalisation layers, shown in \autoref{fig:mlpgnngatplusarchitecture}, were added to normalise the output of each layer with the mean and the standard deviation described in \citet{ba_layer_2016} to improve training stability by preventing vanishing or exploding gradients. \autoref{fig:mlpgnngatplusarchitecture} also shows two residual connections, which act to stabilise loss updates and preserve graph metric fidelity during training by allowing direct information flow from layer 1 $\rightarrow$ 3 and layer 2 $\rightarrow$ 4. Stabilisation occurs because the model can maintain salient local information, as it has access to pre-aggregated node features \citep{huang_residual_2019}. Finally, we implemented a decaying learning rate scheduler that reduced the learning rate by $30\%$ when the validation loss started to plateau. The reduction in the learning rate resulted in a less oscillatory (but more stagnant) loss curve after $\sim4,500$ out of the total $10,000$ epochs as can be seen in \autoref{fig:gat_4H_plus_10_9_loss_accuracy_curves}. The regularisation procedures recommended by recent work combined with a 4-headed GAT layer resulted in our GAT$+$ model, \autoref{tab:resultssummary} also shows the performance of the GAT model, with a single attention head and without any regularisations.

\begin{table}
    \centering

    \begin{tabular}{|l|c|c|c|c|} 
        \hline
        Environment & Weight ($10^{9}\,M_{\odot}$) & Weight ($10^{10}\,M_{\odot}$) \\
         \hline
        Void   & 6.59 & 12.10 \\ \hline
        Wall   & 1.03 & 1.20 \\ \hline
        Filament   & 0.55 & 0.52 \\ \hline
        Cluster  & 0.93 & 0.86 \\
        \hline
    \end{tabular}
    \caption{Class weights used in the loss function for the two stellar mass cuts. The weights are computed from the inverse class frequencies, including galaxies within the buffer region between environments. The $10^{9}\,\mathrm{M}_{\odot}$ mass-cut (left column) is the primary case used for the results presented in this work, as it provides a larger and more representative sample across all cosmic web environments.}
    \label{tab:environmentweights}
\end{table}

\subsection{Performance summary on the test split}
\label{sec:perfsummary}

\begin{figure}
    \centering
    \includegraphics[width=\linewidth]{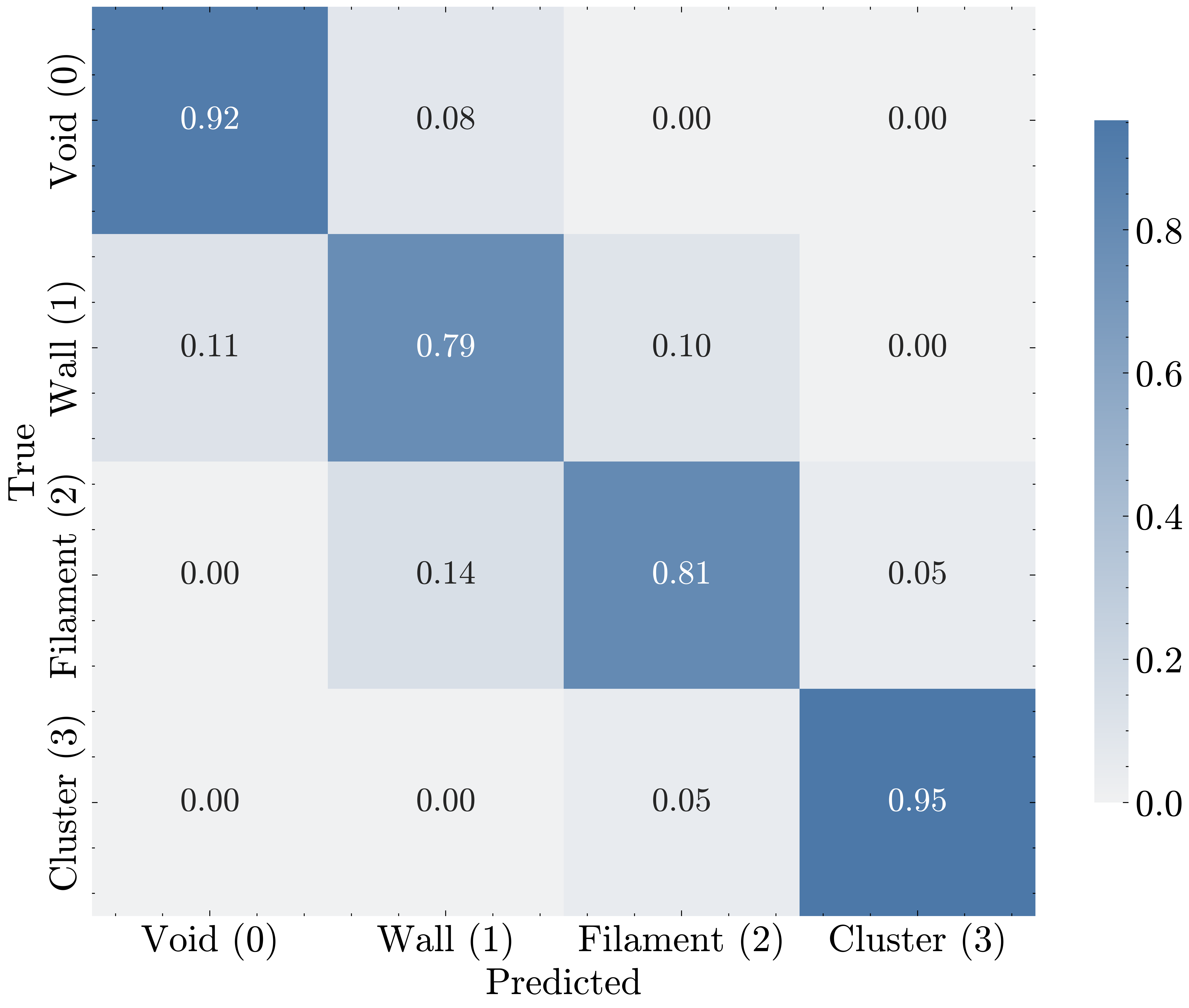}
    \caption{Confusion matrix for the GAT+ model. Results are shown for the $\mathrm{10^{9}\,M_{\odot}}$ stellar mass cut, comparing model predictions (x-axis) against the true T-WEB environment labels (y-axis). The blue colour scale indicates the normalised density of predictions in each class. The matrix illustrates that the model performs best for filament and cluster classifications, with the most confused environments being walls and filaments.}
    \label{fig:gat_4H_plus_10_9_confusion}
\end{figure}

\autoref{tab:resultssummary} summarises the performance of the test-set for all models. The GAT+ architecture achieves the highest overall accuracy (85\%), significantly outperforming the GCN (69-70\%) and the MLP (68\%). The normalised confusion matrix (\autoref{fig:gat_4H_plus_10_9_confusion}) shows that the vast majority of galaxies are correctly classified in all environments. The darkest diagonal elements correspond to voids and clusters, each exceeding $90\%$ accuracy, whereas walls and filaments remain the most confounded classes, with $\sim10$-$14\%$ of galaxies misclassified between the two. This confusion is physically expected because walls and filaments occupy intermediate, overlapping density regimes and share multi-scale substructures, whereas voids and clusters represent the most distinct environments. Tables of the precision, recall, and F1 performance metrics for all the models and their definitions are provided in \autoref{appendix:additionalresults}. The improved performance of the GAT+ model over the GCN model likely arises from its attention mechanism, which allows the network to assign higher weights to informative neighbouring galaxies and down-weight noisy or connections that cross environment boundaries. We also note that both the GCN and GAT model were implemented with Nvidia CUDA acceleration for PyTorch\footnotemark[3] to minimise the training time.

\footnotetext[3]{\url{https://docs.pytorch.org/docs/stable/index.html/}}

\begin{table}
    \centering
    \begin{tabular}{|c|c|c|}
        \hline
	Algorithm &  Accuracy & Training Time (min) \\ \hline
        Random Forest$^{*}$ & 65\% & 2 \\ \hline
        XGBoost$^{*}$ & 70\% & 3 \\ \hline
        Random Forest & 71\% & 4 \\ \hline
        XGBoost & 72\% & 4 \\ \hline
        MLP &  68\% & 5  \\ \hline
        GCN & 69\% & 30 \\ \hline
        GAT & 75\% & 30 \\ \hline
        \textbf{GAT+} & \textbf{85\%} & 30\\ \hline
    \end{tabular}
    \caption{Summary of model accuracies and training times for cosmic web classification. The table lists the validation accuracy and approximate training time for each algorithm. Starred ($^{*}$) rows indicate reference results from \citet{tsizh_large-scale_2020}, included for comparison. The GAT+ model refers to the final model used in this work, incorporating multi-head attention.}
    \label{tab:resultssummary}
\end{table}

\subsection{Feature importance and mutual information}
\label{sec:feature_importance}

To understand which graph metrics are most informative for classification, we examine the statistical association between each input feature and the true cosmic web environments using mutual information \citep{kraskov_estimating_2004, ross_mutual_2014}. Mutual information is useful because it captures non-linear and non-monotonic statistical dependencies.

\autoref{fig:MI} shows that the clustering coefficient exhibits the strongest statistical association with the cosmic web environment. This metric encodes the local substructure in the Delaunay graph by measuring the extent to which nodes cluster together \citep{de_regt_network_2018}. High clustering coefficients correspond to dense, multiply connected regions typical of filaments and clusters, whereas low coefficients indicate sparse, anisotropic configurations characteristic of walls and voids. Interestingly, \citet{tsizh_large-scale_2020} found the opposite trend, identifying the clustering coefficient as the least informative feature. This discrepancy arises from the use of a proximity-based graph with a fixed linking length, which truncates the number of possible triangles in dense regions. In contrast, the Delaunay algorithm guarantees a full space-filling triangulation, ensuring that all relevant local connectivity is captured. The separation of environments in the mean edge length distribution (\autoref{fig:meanel}), and other distributions of graph metrics not shown, is consistent with these trends.

Metrics such as minimum and maximum edge lengths, and node degree show the weakest associations, suggesting that they vary relatively little across environments. This was confirmed by their poorly separated distributions. However, the filament population exhibits a noticeable tail toward smaller minimum edge lengths, likely reflecting its multi-scale substructure: filaments contain dense cores approaching cluster-like densities and tenuous outer regions that bridge towards walls and voids \citep{aragon-calvo_multiscale_2010, bahe_galaxies_2025}. This implies that the minimum edge length is particularly responsive to embedded small-scale density contrasts, while the density, neighbour density, and shape eigenvalue metrics trace smoother, more global variations.

\subsection{Latent structure and learnt representations}
\label{sec:latent_structure}

\begin{figure*}
    \centering
    \includegraphics[width=\linewidth]{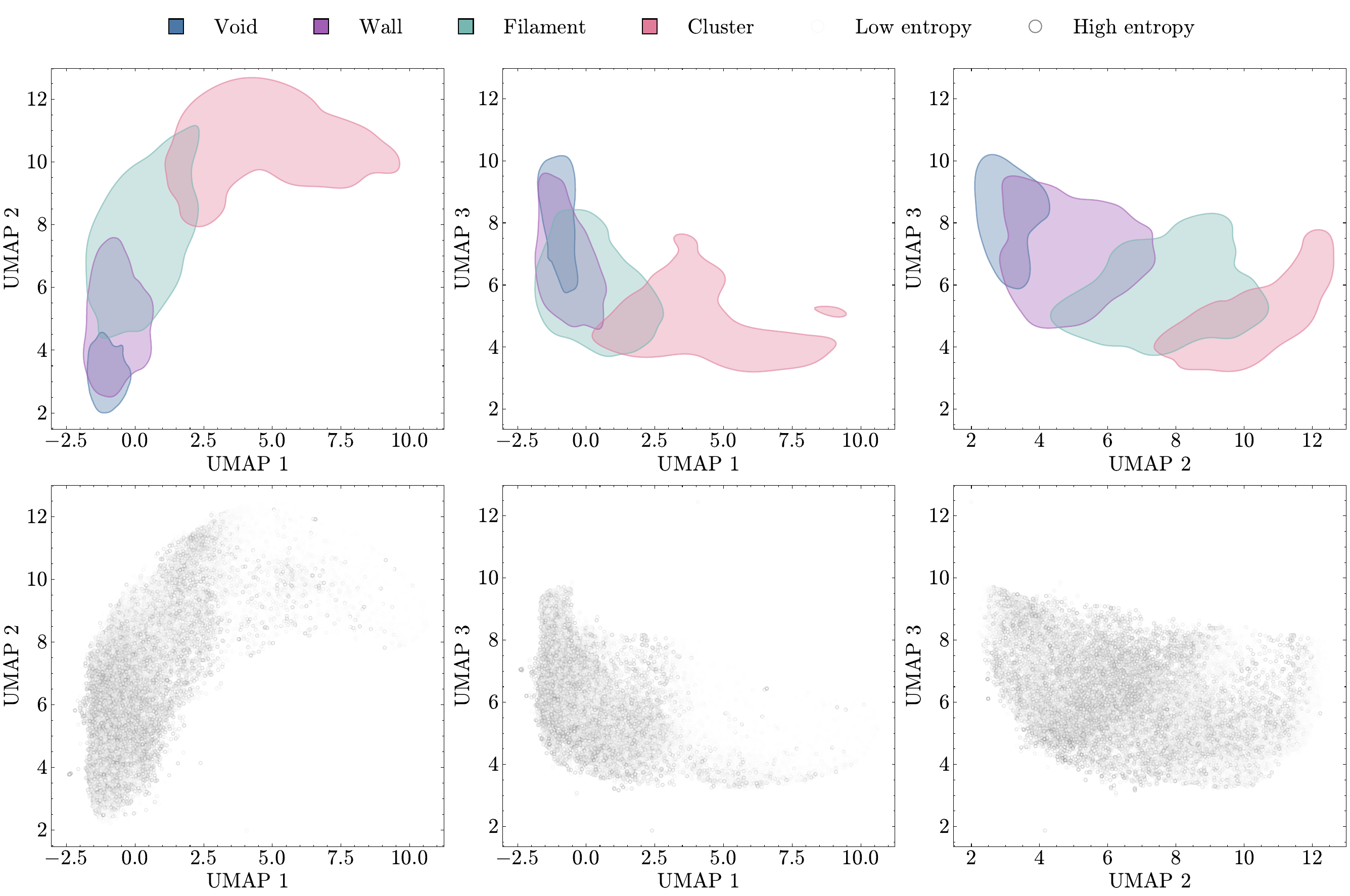}
    \caption{UMAP projections into three-dimensions represented by the columns. Top row: UMAP contours of the GAT+ embeddings, but only limited to the test galaxy set and coloured by the model-predicted environments. Bottom row: GAT+ embeddings of the test galaxy set in UMAP projected space. Darker edges represent higher model uncertainty, which is derived from the Shannon entropy of the output probabilities. The overlap of cosmic web environments in the top figure corresponds with regions of high model uncertainty given by the bottom figure, indicating that the GAT+ model learns a physically meaningful representation of the cosmic web.}
    \label{fig:UMAP}
\end{figure*}

To visualise the separability of the cosmic web environments from the graph metrics, we applied UMAP \citep{mcinnes_umap_2020} dimensionality reduction to the input graph metrics (not shown here) and the learnt GAT+ embeddings, which we coloured by the true cosmic web environments and the GAT+ predictions respectively. The UMAP projection of the ten input features forms overlapping but weakly separable manifolds or clusters. This is reflected in a near-zero silhouette score ($-0.001$). A silhouette score measures how well each data point is assigned to its class with scores ranging from $[-1,+1]$, where higher scores indicate better-defined classes. By contrast, as seen in \autoref{fig:UMAP}, the UMAP of the GAT+ embeddings shows a more distinct clustering by environment, with a silhouette score of $0.108$. This indicates that the network learnt a representation that better aligns with the physical structure of the cosmic web. When only the test split galaxies are included and then coloured by the GAT+ predicted (see top row of \autoref{fig:UMAP}), the embeddings yield an even higher silhouette score ($0.161$), suggesting that the model generalises this separation to unseen data.

In addition, we examined the predictive entropy of the GAT+ model to quantify the confidence of classification, which we calculated using the Shannon entropy $H$ \citep[see][]{shannon_mathematical_1948}. This is defined in terms of the output Softmax probabilities of the GAT+ model for each cosmic web environment, $p_i, i \in [0,3]$ corresponding to $\mathrm{\{Void, Wall, Filament, Cluster\}}$, respectively.

\begin{equation}
    H = - \sum_{i = 0}^{3} p_i\, \mathrm{log}\, p_i.
    \label{eq:shannon_entropy}
\end{equation}

Galaxies with flat output probabilities across environments (e.g. $p = [0.25, 0.25, 0.25, 0.25]$) correspond to high entropy ($H \approx 1$), indicating low confidence, while sharply peaked probabilities ($H \approx 0$) denote confident classifications. The distribution of entropies from the GAT+ model peaked between $0.5-1$, implying that a large fraction of galaxies occupy regions of intrinsic uncertainty, such as void-wall or wall-filament transitions. This pattern reflects the continuous nature of the cosmic web. The bottom row of \autoref{fig:UMAP} shows the test set galaxies in UMAP-projected space and is represented by unfilled circles. The weight of the edges of these circles is determined by the Shannon entropy (\autoref{eq:shannon_entropy}). Galaxies in the transitional regions exhibit darker edges, highlighting that the model uncertainty encodes physically meaningful ambiguity.

\section{Discussion and conclusions}
\label{sec:Discussion}

In this investigation, we developed a three-step methodology using graph neural network models for identifying the dark matter cosmic web environments of a mass-selected catalogue of galaxies ($M_{\star} > 10^{9}\,\mathrm{M}_{\odot}$) from the IllustrisTNG-300 simulation. First, we assigned a T-Web derived environment to each galaxy based on its position. Second, we constructed a Delaunay triangulation graph to capture the relational structure of the distribution of galaxies and summarised this with a set of ten graph metrics associated to each galaxy (see \autoref{tab:graph_metrics}). Third, we trained the GAT$+$ model to learn the relationship between each galaxy's graph metrics and its assigned T-Web environment. When applied to a test data set, the GAT$+$ model classified the dark matter cosmic web environments of galaxies with an accuracy of $85\%$, representing a significant improvement over the non-neural network, MLP, and GCN models that we tested.

\subsection{Comparison of model architectures}
\label{sec:model_comparison}

For context, we first compare our models to the classical machine-learning algorithms implemented by \citet{tsizh_large-scale_2020}. They applied random forest and XGBoost classifiers to a complex graph with a fixed linking length of $2.4\,\mathrm{h^{-1}\,Mpc}$, achieving accuracies of $65\,\%$ and $70\,\%$ respectively. In our implementation using the Delaunay graph and class-weighted loss functions (see \autoref{tab:environmentweights}), the  algorithms improve modestly: the random forest reaches $71\,\%$ and XGBoost $72\,\%$ (\autoref{tab:resultssummary}). These results indicate that the graph construction itself captures valuable structural information. By triangulating the distribution of galaxies rather than imposing a fixed linking scale, the Delaunay graph is a richer description of the influence of the dark matter cosmic web on the galaxy distribution.

Despite this improvement, tree-based methods remain limited by their piecewise-constant decision boundaries. They partition the feature space through sequential axis-aligned splits, which cannot easily capture the smooth, high-order interactions among geometric descriptors that describe the cosmic web. This is evident in the poor recall scores in \autoref{tab:RandomForest_precision_recall_f1} and \autoref{tab:xgboost_precision_recall_f1}. Deep-learning models are better suited to approximate such non-linear boundaries.

The baseline MLP achieves an accuracy similar ($68\,\%$) to the random forest and XGBoost algorithms, despite being more expressive because it treats every galaxy independently and, therefore, ignores the graph connectivity that physically links nearby galaxies. Despite a marginal improvement in overall accuracy, MLP produces a more balanced classification (see \autoref{tab:MLP_precision_recall_f1}). Introducing structural context and neighbourhood feature aggregation in the GCN raises the accuracy slightly to $69$-$70\,\%$, demonstrating that it adds some discriminative power. However, the improvement is small because the ten graph metrics already encode first-order neighbourhood geometry; uniform aggregation in the GCN adds little beyond this.

The GAT and GAT$+$ architectures deliver the largest performance gains. A single-head GAT improves to $75\,\%$, and the GAT$+$ model—with four attention heads, residual connections, layer normalisation, and a dynamic learning-rate scheduler—achieves $85\,\%$. The attention mechanism allows the network to weight each neighbour adaptively, learning which connections are physically informative and which should be suppressed. Multi-head attention captures multiple relational patterns in parallel (for instance, dense cluster cores versus filament spines), while residual connections and normalisation layers stabilise training and mitigate the over-smoothing that affects deeper GCNs. Repeated GCN layers have been shown to lead to a convergence of node features \citep{li_deeper_2018, huang_residual_2019, wei_opnet_2024}, resulting in the GCN losing local discrimination power. This is particularly problematic in transitional regions between voids, walls, filaments, and clusters. This tendency can be compounded by the GCN's weaknesses with slightly heterophilous graphs \citep{ma_few-shot_2021}. Investigating whether this is a significant factor is beyond the scope of this investigation. However, a potential future direction could use metrics proposed by \citet{chen_measuring_2019} to measure over-smoothing.

The improvement from GCN to GAT$+$ demonstrates that attention mechanisms can disentangle heterogeneous local environments within the slightly heterophilous Delaunay graph. In physical terms, the model learns to emphasise coherent, homophilous connections, such as galaxies along the same filament, while down-weighting links across boundaries, for example, between walls and filaments. This is supported by the clearer separation of the GAT$+$ embeddings in \autoref{fig:UMAP}.

\subsection{Continuous environments and classification limits}
\label{sec:continuous_environments}

The residual confusion between walls and filaments (shown by \autoref{fig:gat_4H_plus_10_9_confusion} and the high entropies in \autoref{fig:UMAP}) underscores that the cosmic web is inherently continuous. The finite threshold $\lambda_{\text{th}}$ \citep{foreroromero_dynamical_2009} required to distinguish between environments ultimately reflects how many regions exist in intermediate states of collapse, producing physically fuzzy boundaries, reflected in the overlapping UMAP manifolds (\autoref{fig:UMAP}).

A promising future direction is to reformulate the problem as a regression task, predicting the continuous deformation tensor eigenvalues (or their normalised collapse fractions) rather than discrete classes. This would yield a probabilistic, continuous mapping of the cosmic web that more faithfully captures its physical structure and timescale dependence. 

\subsection{Advantages of graph-based learning}
\label{sec:advantages}

Our results highlight several advantages of graph-based learning for cosmic web studies. By encoding the galaxy distribution into graph connectivity, the method bypasses the need to construct a smoothed density field by learning directly from the discrete data. Graph representations naturally capture the geometry and topology of the cosmic web (e.g. connecting nodes to delineate filaments, walls, clusters, and voids), providing an interpretable framework for large-scale structure. Moreover, the flexibility to choose graph metrics that capture local geometries, combined with the structure-aware architecture of GNNs allows our method to effectively capture a fuller dynamic range of cosmic web structure, yielding physically motivated classifications for galaxies. However, our method is limited by the assumptions of the training simulation. This means that we are limited to the domains we can apply our current model for inference, where a dark matter field cannot be sampled or where a complete sample of galaxies is not available. Domain shifts are an active area of research.

\subsection{Application to observational data}
\label{sec:application}

Exploiting the newly available catalogues of galaxy spectra to build a statistically robust and multi-scale understanding of the role of cosmic web environments in galaxy evolution is now of active interest. However, it should be noted that the analysis presented in this investigation was performed on an idealised dataset, which does not include important observational effects and survey systematics. Applying this framework to real survey data, such as DESI, will require domain adaptation. This is achieved by training on simulations that model survey geometry, spectroscopic completeness, fibre incompleteness, sample selections, and redshift-space distortions. This is an essential next step before deployment on large-scale galaxy surveys.

Once adapted, our method can provide a probabilistic cosmic web context for galaxy surveys. We plan to apply our method to infer the probabilities for each cosmic web environment of galaxies in the ongoing DESI survey. This will enable environmental classification on an unprecedented scale.

The continuing development of graph neural networks combined with established Bayesian methods for uncertainty will further enhance our ability to characterise the cosmic web, bridging simulations and observations in the new era of precision extragalactic astronomy.

\section*{Acknowledgements}

This work has been enabled by the STFC-supported UCL Centre for Doctoral Training in Data Intensive Science. DK thanks Lucas Makinen and Niall Jeffrey for their support and guidance. KN acknowledges support from the Royal Society grant number URF\textbackslash R\textbackslash 231006.

\section*{Data Availability}

All data used in this analysis are produced from publicly available datasets and software packages. The IllustrisTNG-300 \citep{nelson_illustristng_2021} simulation data can be downloaded from the TNG collaboration data access page\footnotemark[4].

\footnotetext[4]{\url{https://www.tng-project.org/data/}}.


\bibliographystyle{rasti}
\bibliography{references}



\appendix

\section{Graph metric definitions}
\label{appendix:graph_metric_defn}
To summarise the local geometric and connectivity of the Delaunay graph, we chose ten node-level graph metrics. We define the following graph metrics in terms of the weighted adjacency matrix $\mathbfss{A} \in \mathbb{R}^{n \times n}$, where the components $w_{ij}$ are the lengths of the edges between two nodes, or 0 if there is no edge (see \autoref{eq:adjacencyelement}). We also define the set of neighbours of a given node $i$ as $\mathcal{N}(i) = \{ j | w_{ij} > 0 \}$.

\begin{itemize}

\item Degree: This measures the number of edges connected to each node in the Delaunay graph, and for weighted graphs, the degree is then the sum of edge lengths.

\begin{equation}
   d_i = \sum_{j \in \mathcal{N}(i)} w_{ij}
\end{equation}

\item Clustering \citep[see][]{onnela_intensity_2005}: This measures the tendency of nodes to form tightly connected neighbourhoods. For weighted graphs, the clustering coefficient of a node is defined in terms of the lengths of the connecting edges normalised by the longest edge length, $\hat{w}_{ij} = \frac{w_{ij}}{\text{max}(w)}$. 

\begin{equation}
    c_i = \frac{1}{d_i (d_i -1)} \sum_{j,k \in \mathcal{N}(i)} (\hat{w}_{ij}\hat{w}_{ik}\hat{w}_{jk})^{\frac{1}{3}},
\end{equation}

\item Mean edge length: This is the arithmetic mean of the edge lengths connected to a given node.

\begin{equation}
    \bar{e}_i = \frac{1}{d_i} \sum_{j \in \mathcal{N}(i)} w_{ij}.
\end{equation}

\item Minimum edge length: The minimum edge length of the edges that connect to a given node.

\begin{equation}
    e_i^{(min)} = \text{min}_{j\in\mathcal{N}(i)}(w_{ij})
\end{equation}

\item Maximum edge length: The maximum edge length of the edges that connect to a given node.

\begin{equation}
    e_i^{(max)} = \text{max}_{j\in\mathcal{N}(i)}(w_{ij})
\end{equation}

\noindent For densities from the Delaunay graph, we consider the volume of the tetrahedra (a simplex in three dimensions) that connect to a given node. We define $\mathcal{T}_{i}$ as the number of Delaunay tetrahedra that include node $i$, and $\sum_{t\in \mathcal{T}_i} V(t)$ is the total volume of all tetrahedra that include node $i$. \newline

\item The tetrahedral density (tetra dens in \autoref{tab:graph_metrics}) is the weighted average of the inverse of the total volume of tetrahedra that includes a given node.

\begin{equation}
    \rho_i = \frac{\mathcal{T}_{i}}{d_i \sum_{t\in \mathcal{T}_i} V(t)}.
\end{equation}

\item Neighbouring tetrahedra density (neigh tetra dens in \autoref{tab:graph_metrics}): This is defined as the arithmetic mean of a given node's tetrahedral density over its neighbours.

\begin{equation}
    \rho_{\mathcal{N}(i)} = \frac{1}{d_i} \sum_{j \in \mathcal{N}(i)} \rho_j.
\end{equation}

\noindent The final three graph metrics quantify the shape of the local point cloud around each node by measuring the anisotropy of its neighbouring galaxies. This can be expressed in a coordinate-independent manner using the eigenvalues of the covariance matrix of neighbour positions. Let the three-dimensional position of each node or galaxy be $\mathbfit{X}_i \in \mathbb{R}^3$. For a node $i$ with degree $d_i$, the neighbourhood centre of mass is

\begin{equation}
    \bar{\mathbfit{X}}_i = \frac{1}{d_i} \sum_{j \in \mathcal{N}(i)} \mathbfit{X}_j.
\end{equation}

\noindent Using this, we compute the covariance matrix of the neighbouring positions:

\begin{equation}
    \mathbfss{C}_{i} = \frac{1}{d_i} \sum_{j \in \mathcal{N}(i)} (\mathbfit{X}_j - \bar{\mathbfit{X}}_{i}) (\mathbfit{X}_j - \bar{\mathbfit{X}}_i)^{\mathbb{T}}.
\end{equation}

\item The eigenvalues of $\mathbfss{C}_{i}$ can be ordered such that $I_1(i) \leq I_2(i) \leq I_3(i)$. These shape eigenvalues characterise the local morphology:

\begin{itemize}
    \item $I_1 \approx I_2 \approx I_3 \Rightarrow$ isotropic or spherical distribution of neighbours, typical of cluster-like regions or uniformly sparse void regions,
    \item $I_1 \ll I_2 \approx I_3 \Rightarrow$ flattened, planar configuration, typical of wall-like environments,
    \item $I_1 \approx I_2 \ll I_3 \Rightarrow$ elongated configuration aligned along a preferred axis, typical of filamentary environments.
\end{itemize}

\end{itemize}

\section{Additional Results}
\label{appendix:additionalresults}

In this section, we present the full precision, recall, and F1 scores for each cosmic web environment predicted by the MLP and GAT+ models, as well as the reference random forest and boosted decision tree models referenced in \autoref{tab:resultssummary}. While overall classification accuracies provide a high-level summary of model performance, these class-resolved metrics allow for a more detailed evaluation of how well each architecture distinguishes voids, walls, filaments, and clusters. Precision and recall highlight different aspects of the classification behaviour - contamination versus completeness - and the F1 score provides a balanced harmonic mean of the two. Therefore, these tables illustrate the relative strengths of each model and the general pattern of confusion across the four environments, especially between walls and filaments. All scores are computed using the 

\begin{table}
  \centering
  \begin{tabular}{|c|c|c|c|}
    \hline
              & Precision & Recall   & F1-Score \\ \hline
    Void (0)  & 0.288618  & 0.696078 & 0.408046 \\ \hline
    Wall (1)  & 0.533570  & 0.652416 & 0.587038 \\ \hline
    Filament (2) & 0.754267  & 0.602435 & 0.669855 \\ \hline
    Cluster (3)  & 0.750050  & 0.810236 & 0.778982 \\ \hline
    Accuracy  & 0.675562  & 0.675562 & 0.675562 \\ \hline
    Macro Avg & 0.581626  & 0.690292 & 0.610980 \\ \hline
    Weighted Avg & 0.698580 & 0.675562 & 0.679774 \\ \hline
  \end{tabular}
  \caption{Class-resolved performance metrics (precision, recall, F1 score) for the MLP model. These values correspond to the test set for galaxies with stellar mass $ > \,10^9 \mathrm{M}_{\odot}$, split by T-Web environment. The table highlights the baseline performance of a non-graph neural network model using the node-level graph metrics defined in Appendix~\ref{appendix:graph_metric_defn}.}
  \label{tab:MLP_precision_recall_f1}
\end{table}

\begin{table}
  \centering
  \begin{tabular}{|c|c|c|c|}
    \hline
    & Precision & Recall & F1-score \\ \hline
    Void (0) & 0.625000 & 0.212418 & 0.317073 \\ \hline
    Wall (1) & 0.629535 & 0.607497 & 0.618319 \\ \hline
    Filament (2) & 0.713353 & 0.766411 & 0.738931 \\ \hline
    Cluster (3) & 0.802475 & 0.759488 & 0.780390 \\ \hline
    Accuracy & 0.720909 & 0.720909 & 0.720909 \\ \hline
    Macro avg & 0.692591 & 0.586454 & 0.613678 \\ \hline
    Weighted avg & 0.720572 & 0.720909 & 0.718088 \\ \hline
  \end{tabular}
  \caption{Class-resolved performance metrics for the optimised random forest model. Setting $\texttt{n\_estimators}=500$, $\texttt{max\_depth}=50$, $\texttt{min\_samples\_split}=2$, $\texttt{min\_samples\_leaf}=1$, $\texttt{bootstrap}=\texttt{True}$ resulted in the highest classification accuracy for the test set.}
  \label{tab:RandomForest_precision_recall_f1}
\end{table}

\begin{table}
  \centering
  \begin{tabular}{|l|c|c|c|}
    \hline
    & Precision & Recall & F1-score \\ \hline
    Void (0) & 0.585938 & 0.245098 & 0.345622 \\ \hline
    Wall (1) & 0.651525 & 0.608736 & 0.629404 \\ \hline
    Filament (2) & 0.719892 & 0.774484 & 0.746191 \\ \hline
    Cluster (3) & 0.801039 & 0.769247 & 0.784821 \\ \hline
    Accuracy & 0.728552 & 0.728552 & 0.728552 \\ \hline
    Macro avg & 0.689598 & 0.599391 & 0.626510 \\ \hline
    Weighted avg & 0.727056 & 0.728552 & 0.725718 \\ \hline
  \end{tabular}
  \caption{Class-resolved performance metrics for the optimised XGBoost model. Setting $\texttt{n\_estimators}=500$, $\texttt{max\_depth}=8$, $\texttt{learning\_rate}=0.2$, $\texttt{colsample\_bytree}=0.7$, $\texttt{min\_child\_weight}=1$, and $\texttt{gamma}=0$ resulted in the highest classification accuracy for the test set.}
  \label{tab:xgboost_precision_recall_f1}
\end{table}

\begin{table}
  \centering
  \begin{tabular}{|l|c|c|c|}
    \hline
    & Precision & Recall & F1-Score \\ \hline
    Void (0) & 0.609462 & 0.878342 & 0.719606 \\ \hline
    Wall (1) & 0.774295 & 0.838859 & 0.805285 \\ \hline
    Filament (2) & 0.928106 & 0.842309 & 0.883129 \\ \hline
    Cluster (3) & 0.941608 & 0.960928 & 0.951170 \\ \hline
    accuracy & 0.875006 & 0.875006 & 0.875006 \\ \hline
    macro avg & 0.813368 & 0.880109 & 0.839797 \\ \hline
    weighted avg & 0.883121 & 0.875006 & 0.876907 \\ \hline
  \end{tabular}
  \caption{Class-resolved performance metrics for the GCN model. These values correspond to the test set of galaxies with a stellar mass $>\,10\,\mathrm{M}_{\odot}$ split by T-Web environment. This table highlights the marginal improvements of the GCN model over the MLP due to its fixed averaging of neighbouring node features. Architectural details are provided in \autoref{fig:mlpgnngatplusarchitecture}.}
  \label{tab:GCN_precision_recall_f1}
\end{table}

\begin{table}
  \centering
  \begin{tabular}{|l|c|c|c|}
    \hline
    & Precision & Recall & F1-Score \\ \hline
    Void (0) & 0.541930 & 0.915775 & 0.680915 \\ \hline
    Wall (1) & 0.743020 & 0.786016 & 0.763913 \\ \hline
    Filament (2) & 0.908611 & 0.814361 & 0.858908 \\ \hline
    Cluster (3) & 0.926292 & 0.952899 & 0.939407 \\ \hline
    accuracy & 0.848834 & 0.848834 & 0.848834 \\ \hline
    macro avg & 0.779963 & 0.867263 & 0.810786 \\ \hline
    weighted avg & 0.860178 & 0.848834 & 0.851409 \\ \hline
  \end{tabular}
  \caption{Class-resolved performance metrics for the GAT+ model. The GAT+ architecture incorporates multi-headed attention, residual connections, and layer normalisation. These enhancements yield the highest precision and recall across all environments, particularly improving the separation between walls and filaments.}
  \label{tab:GATplus_precision_recall_f1}
\end{table}


\bsp	
\label{lastpage}
\end{document}